\documentclass[letterpaper,twocolumn,10pt]{article}
\usepackage{usenix-2020-09}

\usepackage{tikz}
\usepackage{amsmath}
\usepackage{amsfonts}
\usepackage{xspace}
\usepackage{xurl}
\usepackage{tabularx}
\usepackage{subcaption}

\newcommand*{\sn}{\textsc{ZK-IMG}\xspace}
\newcommand*{\halo}{\texttt{halo2}\xspace}

\newcommand{\minihead}[1]{{\vspace{.45em}\noindent\textbf{#1.} }}

\begin{document}

\date{}

\title{ZK-IMG: Attested Images via Zero-Knowledge Proofs to Fight Disinformation}

\author{
{Daniel Kang}\\
UIUC
\and
{Tatsunori Hashimoto}\\
Stanford
\and
{Ion Stoica}\\
Berkeley
\and
{Yi Sun}\\
University of Chicago
} 

\maketitle

\begin{abstract}

Over the past few years, AI methods of generating images have been increasing in
capabilities, with recent breakthroughs enabling high-resolution, photorealistic
``deepfakes'' (artificially generated images with the purpose of misinformation
or harm). The rise of deepfakes has potential for social disruption. Recent work
has proposed using ZK-SNARKs (zero-knowledge succinct non-interactive argument
of knowledge) and attested cameras to verify that images were taken by a camera.
ZK-SNARKs allow verification of image transformations non-interactively (i.e.,
post-hoc) with \emph{only} standard cryptographic hardness assumptions.
Unfortunately, this work does not preserve input privacy, is impractically slow
(working only on 128$\times$128 images), and/or requires custom cryptographic
arguments.

To address these issues, we present \sn, a library for attesting to image
transformations while hiding the pre-transformed image. \sn allows application
developers to specify high level image transformations. Then, \sn will
transparently compile these specifications to ZK-SNARKs. To hide the input or
output images, \sn will compute the hash of the images inside the ZK-SNARK.  We
further propose methods of chaining image transformations securely and
privately, which allows for arbitrarily many transformations. By combining these
optimizations, \sn is the first system to be able to transform HD images on
commodity hardware, securely and privately.

\end{abstract}

\section{Introduction}

In recent years, \emph{deepfakes} have been proliferating. Deepfakes are
``videos and images that have been digitally manipulated to depict people saying
and doing things that never happened'' \cite{patterson2019from}. One common use
of deepfakes is to fool viewers for the purpose of misinformation, typically on
social media \cite{patterson2019from}. They have also been used to fool
companies into sending money to hackers \cite{law2022better} and fake personas
to seed misinformation for business scams \cite{vincent2022binance}. Their
ability to fool and prevalence has been driven by artificial intelligence
methods \cite{ramesh2022hierarchical, rombach2022high}.

The rise of deepfakes raises a critical question: how can we verify the
authenticity of visual media in the face of malicious adversaries? Furthermore,
many applications have additional desiderata. First, the image from the camera
should be able to be \emph{transformed} by a series of permissible transforms
from the source image. For example, an image should be able to be cropped or
selectively blurred to exclude sensitive information. Second, the visual media
should be verifiable \emph{non-interactively}, or after the image has been taken
and transformed. Non-interactivity is particularly important for social media as
social media providers and consumers are not present at the time of image
generation. Third, arbitrary third parties should be able to verify the
transformed image.

One method to attest to transformed images to is to combine attested cameras
(which sign images immediately on being taken) with cryptographic techniques to
verify the transformations. In particular, ZK-SNARKs (zero-knowledge succinct
non-interactive argument of knowledge) allow arbitrary computations to be
verified without revealing any information about the inputs or intermediate
steps of the computation. Furthermore, Zk-SNARKs are non-interactive and produce
certificates that any third party can verify. 

Unfortunately, prior work for ZK-SNARKs for image edits are impractical for
several reasons \cite{naveh2016photoproof, ko2021efficient, datta2022using}.
First, \emph{they require revealing the original or intermediate images to
verify the transformations} \cite{naveh2016photoproof, datta2022using}. If the
original image could be released, then there is no need for private image
transformations as the verifier can replay the transformations. Although these
protocols could be modified to address this shortcoming, we show that doing so
can add up to 20$\times$ overheads. Second, they operate on only images that are
impractically small (128$\times$128 or smaller) \cite{naveh2016photoproof,
datta2022using}. Third, they require custom arguments for specific image
transformations \cite{naveh2016photoproof, ko2021efficient}.  For example, some
prior work can only perform image crops \cite{ko2021efficient}.


As a first step towards verification of visual media, we present \sn, the first
system to attest the validity of \emph{arbitrary and arbitrarily many} image
transforms on HD (720p) images on commodity hardware. \sn takes as input a
camera-attested image and a high-level specification of the transformations to
apply. It then produces a ZK-SNARK proof of the transformations and the output
image. \sn can attest transformations of HD images on commodity hardware, which
is two orders of magnitude larger than prior work.

To produce ZK-SNARK proofs, \sn leverages recent \emph{general-purpose} ZK-SNARK
proving systems, specifically \texttt{halo2} \cite{halo2}. We demonstrate how to
efficiently implement a wide range of image transforms in \texttt{halo2},
including ``physical'' transformations (crops, rotations, flips, translations,
resizes), colorspace conversions (RGB to YCbCr and YCbCr to RGB), filters
(sharpen, blur), and other standard operations (white balance, contrast
adjustment). We implement these efficiently in \sn by fusing operations and
packing computations efficiently in the ZK-SNARK construction.

In addition to implementing individual image operations, we also provide an
end-to-end method of chaining together \emph{arbitrarily many} image
transformations without revealing the intermediate outputs. To do so, \sn will
pack as many transformations into a single proof as possible. If not all
transformations fit in a single proof, \sn will split them across proofs. To
avoid revealing information in this process, \sn hashes the inputs and outputs
and reveals only the hashes as part of the proofs. Only the final output is
revealed. By combining these optimizations, \sn is able to verify arbitrarily
many image transforms on HD images on commodity hardware.

We evaluate \sn on a range of image transformations. We show that \sn can
produce proofs for HD image transformations that can be verified in as little as
\emph{5.6 milliseconds} on commodity hardware. Furthermore, \sn can prove
end-to-end image transformations on commodity hardware, costing as little as
\$0.48.

In the remainder of the paper, we provide background on deepfakes in
Section~\ref{sec:background}, use cases in Section~\ref{sec:use-case}, and
ZK-SNARK in Section~\ref{sec:snarks}. We describe \sn's architecture in
Section~\ref{sec:arch} and its detailed implementation in
Sections~\ref{sec:single-image} and \ref{sec:multi-image}. We evaluate \sn in
Section~\ref{sec:eval}. Finally, we discuss related work in
Section~\ref{sec:rel-work} and conclude in Section~\ref{sec:conclusion}.

\section{Background}
\label{sec:background}

In recent years, purposeful spread of fake information and impersonations have
increased in prevalence and capabilities. There are many malicious uses for such
capabilities including spreading misinformation on social media
\cite{allcott2019trends, wang2019systematic}, fooling businesses to wire money
to hackers \cite{stupp2019fraudsters}, and others.  Furthermore, these actors
can range from state actors to ``lone-wolf'' hackers \cite{goldstein2021how,
wakefield2020deepfake}.

\begin{figure}
  \includegraphics[width=\columnwidth]{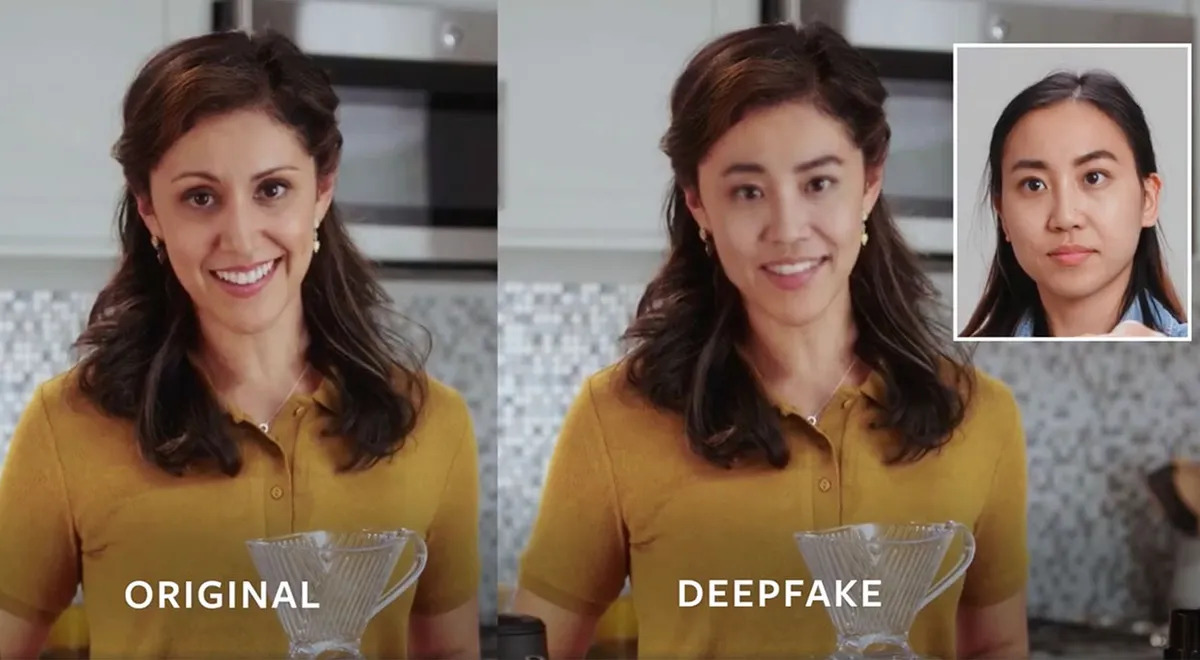}
  \caption{An example of a deepfake \cite{strickland2019facebook}. As shown,
  another person's likeness can be replaced in the image realistically. Although
  this example is innocuous, deepfakes have been used for malicious purposes,
  including theft and the spread of misinformation.}
  \label{fig:deepfake}
\end{figure}

\minihead{Deepfakes}
One way of spreading this misinformation is through the use of deepfakes.
Deepfakes are ``a specific kind of synthetic media where a person in an image or
video is swapped with another person's likeness'' \cite{somers2020deepfakes}. We
show an example in Figure~\ref{fig:deepfake} \cite{strickland2019facebook}.

In recent years, deepfakes have been increasing in their realism. This is driven
in large part due to advances in machine learning (ML) methods, in particular
generative ML methods. For example, models such as Stable Diffusion
\cite{rombach2022high} and others can generate photorealistic images and edit
existing images with high fidelity.

As such, deepfakes have already been used to spread misinformation (e.g., to
fool soldiers into surrendering by faking heads of states)
\cite{wakefield2020deepfake} and steal money (e.g., by impersonating a CEO's
voice) \cite{stupp2019fraudsters}. Finding solutions to deepfakes is an
important and challenging problem.

\minihead{Attested cameras}
One proposal to help address the problem of deepfakes are \emph{attested
cameras}. Attested cameras contain hardware elements that can digitally sign
images \emph{immediately on capture}. Attested cameras further contain hardware
elements to ensure that the camera is tamper-proof and the private key cannot be
extracted without being destroyed.

Given these capabilities, attested cameras can attest that an image was taken by
a particular camera. In this work, we assume the existence of secure (i.e.,
tamper-proof) attested cameras.

\minihead{Using attested cameras}
Given the existence of attested cameras, several organizations (e.g., C2PA) and
prior work has proposed using these cameras to combat deepfakes. In particular,
consumers of digital media can verify that a camera took a particular image
without any assumptions of trust (beyond standard cryptographic assumptions).

However, raw images are rarely released in practice. In practice, nearly all
released images are edited \cite{modems2020edit}. These edits are used to remove
sensitive information, increase the legibility of the image, and for improved
visual quality.

The image edits create a security challenge: the edits must be attested if the
original attested image is not released. Image edits can be attested by software
systems. However, this creates additional surface area for attackers: if a
malicious actor can access the editing software private key, they can sign
arbitrary images. As many users who aim to attest to image edits are
non-experts, this can be unacceptable in high-stakes scenarios. For example,
state actors can compel citizens to give up their private keys, but camera
manufacturers can discard private keys after manufacturing.

In the remainder of this work, we describe several use cases for verified image
transformations and how to attest to image edits in a trustless manner.

\section{Use Cases}
\label{sec:use-case}

Verified image transformations have uses in a range of settings, which have
different application requirements. We describe several example use cases and
their requirements below.

\minihead{Verified images with redaction and editing}
One important use case is to verify an edited image where the original image is
hidden. Verifiers wish to ensure that the images are from an attested camera and
image producers wish to redact and edit the image. For example, a news
organization may wish to redact bystander faces or confidential information. In
the setting of social media, verifiers cannot trust the image producers due to
malicious adversaries. In this setting, the original image cannot be revealed.
Thus, prior work that reveals the original image cannot be used in this setting.

\minihead{Attesting to hidden images}
Another important use case is to attest to an edited image where the edited
image is hidden until some later point in time. To understand why this is
desired, consider the famous ``situation room'' photo of former President Obama
in the white house during Operation Neptune Spear \cite{bowden2012hunt}. Neither
the original image nor the edited image can be revealed at the time the image
was taken. Furthermore, the camera cannot be connected to the internet for
security purposes, which makes trusted third parties (TPPs) for timestamps
infeasible.

Some applications may also desire to attest to a hidden image so that the
hidden, edited image can be used in downstream applications. For example, these
hidden images can subsequently be used as inputs to machine learning models for
biometric identification, where the user wishes to hide the image.

In this setting, the image producer can post a hash and signature of the
original image, a ZK-SNARK of the edit \emph{and a hash of the edited image}.
Namely, the output image is not revealed, only the hash. By publishing the hash
and proof (e.g., on a storage server or blockchain), the image producer can
prove that the image was taken before the output image is revealed.


\section{ZK-SNARKs and Halo2}
\label{sec:snarks}

\subsection{Overview}
In this work, we are interested in verifying function execution. Namely, let $y
= f(x; w)$, where $x$ is the \emph{public input}, $w$ is the \emph{private
input}, and $y$ is the output. In our setting, the public input $x$ is hash of
the pre-transformed image, the private input $w$ is the pre-transformed image
itself, and $y$ is either the hash of the output image or the output image
itself.

ZK-SNARKs allow a \emph{prover} to produce a proof $\pi$ that can convince a
\emph{verifier} of the function execution, where the verifier only has access to
$\pi$, $y$, and $x$ \cite{bitansky2017hunting}. We defer a full description of
ZK-SNARKs and their properties to \cite{bitansky2017hunting}, but highlight
several of its properties.

For the purposes of image transformation attestation, ZK-SNARKs are
\emph{non-interactive} and \emph{zero-knowledge}. Namely, they do not require
communication between the prover and verifier beyond $\pi$, so the proof can be
verified at any point in time. They also reveal no information about $w$ beyond
what is already revealed in $x$ and $y$. As such, they satisfy the criteria we
outlined above.

ZK-SNARKs are also \emph{succinct} (have short and easy to verify proofs),
\emph{complete} (correct proofs will verify successfully), and \emph{knowledge
sound} (computationally bounded provers cannot generate proofs of invalid
executions).

ZK-SNARK proof generation typically involves two steps. The first step,
arithmetization, turns the computation into a system of polynomial equations
over a large finite field that hold if and only if the computation evaluates
correctly. The second step uses a cryptographic proof system (i.e., a
\emph{backend}) to generate the proof from the polynomial equations.

In this work, we use the Halo2 proving system \cite{halo2}, implemented in the
\halo software library. Halo2 uses the Plonkish arithmetization, which contains
several features that are useful for numeric computations beyond previous
proving systems (e.g., Groth16). We now describe the Plonkish arithmetization
and performance considerations with Halo2.

\subsection{Plonkish Arithmetization and Halo2}
We now describe the Plonkish arithmetization. We begin by describing its logical
operations from an application developer perspective, an example of how to
implement a simple circuit, and conclude with performance considerations.

\minihead{Plonkish arithmetization (logical operations)}
We first describe the \emph{logical} operations allowed in the Plonkish
arithmetization from an application developer perspective. In this section, we
use \texttt{numpy} indexing notation where possible.

In the Plonkish arithmetization, the variables of the polynomials take values in
a large prime field and are arranged in a rectangular grid. We denote this grid
as \texttt{c[i][j]}, where \texttt{i} is the row index and \texttt{j} is the
column index.

The first logical operation allowed in the Plonkish arithmetization is a
\emph{copy} operation, which is equivalently an equality operation. The equality
operation enforces that two chosen cells in the grid have equal values, namely
that  \texttt{c[i][j] == c[i'][j']} for chosen indexes \texttt{i}, \texttt{j},
\texttt{i'}, and \texttt{j'}.

The second logical operation allowed in the Plonkish arithmetization is a
\emph{lookup} operation. The lookup operation requires a set of lookup columns,
\texttt{c[:][j1', j2', ..., jn']}. Given these columns, the lookup operation
enforces that the columns in a given row must be in the lookup columns. Namely,
for a row \texttt{i} and a set of columns \texttt{j1}, \texttt{j2}, ...,
\texttt{jn}, the lookup argument enforces \par
\texttt{c[i][j1, ..., jn] = c[i'][j1', ..., jn']}\par
\noindent
for some row index \texttt{i'} in the grid.

The third logical operation allowed in the Plonkish arithmetization is a
\emph{custom gate} operation. The custom gate operation enforces a polynomial
constraint on some configuration of rows and columns in a fixed pattern. Namely,
let \texttt{(a1, b1), (a2, b2), ..., (an, bn)} be offset indexes. Then, for some
polynomial $f$ taking \texttt{n} variables, the custom gate enabled on row
\texttt{i} enforces that\par
$f$(\texttt{c[i + a1][b1]}, ..., \texttt{c[i + an][bn]}) = 0\par
\noindent
In many cases, \texttt{ai} is zero, so the custom gate applies only over a
single row. The application developer may select which rows the custom gate
applies to.

\minihead{Example}
To demonstrate how each operation is used, we provide a toy example. Consider
the setting where a prover wishes to prove that they know values $a$, $b$, and
$c$ such that $a + b = 5$, $a + c = 7$ and $a \in \{0, .., 3\}$. The prover
wishes to keep $a$, $b$, and $c$ hidden.

There are a number of ways to construct an arithmetic circuit satisfying the
properties. As we discuss below, different constructions have performance
implications. We choose a particular construction for the purpose of this
example.

We first lay out the grid as follows\par
{
\centering
\begin{tabular}{l|llll}
    & 0 & 1 & 2 & 3 \\
  \hline
  0 & a & b & 5 & 0 \\
  1 & a & c & 7 & 1 \\
  2 & - & - & - & 2 \\
  3 & - & - & - & 3
\end{tabular}\par
}
\noindent
where the cells with dashes indicate that they can take arbitrary values. In
this construction, we reveal column 2 as public inputs.

We then apply one equality operation, one lookups, and one custom gate:
\begin{enumerate}
  \item We use the equality operator to enforce that \texttt{c[0][0] ==
  c[1][0]}, which constrains the first two cells in the first column to be equal
  to to some value, which we denote $a$.
  \item We use the lookup to enforce that \texttt{c[0][0]} takes a value in
  column 3. This enforces that $a = $ \texttt{c[0][0]} $\in \{0, ..., 3 \}$.
  \item We use a custom gate that enforces \texttt{c[i][0] + c[i][1] - c[i][2] =
  0} (i.e., the row offsets are all 0). We apply this custom gate on rows 0 and
  1. Since \texttt{c[0][2]} and \texttt{c[1][2]} are revealed,
  this enforces that $a + b = 5$ (in row 0) and that $a + b = 7$ (in row 1).
\end{enumerate}

As mentioned, there are a number of ways to construct arithmetic circuits to
satisfy the constraints. As an example, instead of using a single custom gate
with a public column, we could have instead used two custom gates that constrain
\texttt{c[i][0] + c[i][1] - 5 = 0} and \texttt{c[i][0] + c[i][1] - 7 = 0}, and
removed column 2.

Additionally, we could have removed column 3 and instead used a custom gate to
enforce that $a \in \{0, ..., 3\}$. The custom gate would be \texttt{c[i][0] *
(c[i][0] - 1) * (c[i][0] - 2) * (c[i][0] - 3) = 0}. By using this custom gate,
we can also omit rows 2 and 3.

As we discuss below, choices of arithmetization can have dramatic effects on
performance.

\minihead{Halo2}
\halo is a library that implements ZK-SNARK proving via the Plonkish
arithmetization \cite{halo2}. Its ``frontend'' accepts specifications of
Plonkish arithmetic circuits. Given a circuit specification, \halo compiles the
circuit to a ZK-SNARK via a cryptographic ``backend.'' We use a KZG-based
backend, which has small proof sizes (relative to other ZK-SNARK backends)
\cite{kate2010constant}.

\section{Arithmetization Performance Considerations}
We now describe the intuition behind the implementation of the Plonkish logical
operations and its performance implications.

Although an accurate cost model for Plonkish circuits is difficult to construct,
computing relative costs is straightforward. Namely, reducing the number of
rows, columns, permutation arguments, and lookup tables will almost always
reduce the computational costs.  Furthermore, for two circuits with the same
number of permutation arguments, number of lookup tables, and \emph{total} grid
size, the circuit with the smaller number of rows will typically be cheaper.
Finally, for the circuits we consider, the maximum polynomial degree is fixed,
so it is irrelevant for relative cost considerations.

To understand why, we describe several salient properties of Plonkish circuits.
We focus on the grid size, lookup tables, and custom gates.

First, the computational burden of both proving and verification scales with the
number of rows and columns in the grid (although not necessarily linearly in the
total number of cells). As a result, minimizing grid size is imperative.

Second, the number of rows in a valid grid \emph{must be a power of two}. This
constraint is enforced as the proving system operates over subgroups of the
large finite field, which are of size $2^k$ for some $k$. This constraint is
critical. For example, consider a circuit with $2^k + 1$ rows with useful values
(i.e., values that are constrained). Then, the circuit must be size $2^{k + 1}$.
Thus, the number of rows is critical.

Third, in order to implement a lookup, the proving system internally adds an
extra column, two equality constraints (which is internally a ``permutation
argument''), and a low degree custom gate. These additions are in addition to
the lookup column. Furthermore, to enforce the zero knowledge property of the
lookup table, the proving system adds $t$ extra rows to the lookup table (where
$t$ depends on the security parameter of the proving system). This is
particularly salient for image transformations since a lookup table of size
$2^{24}$ actually requires $2^{24} + t$ rows, which forces the grid to have at
least $2^{25}$ rows.

Fourth, the proving system internally adds an extra column for every custom gate
as well. In particular, denote the extra column to be $s_i$. The proving system
replaces the polynomial $f$ with $s \cdot f$. Then, in order to ``select'' a
row, the proving system sets $s_i = 1$, so that $s_i \cdot f = f$. In order to
``deselect'' a row, the proving system sets $s_i = 0$ so that $s_i \cdot f = 0$.

The final setup, proving, and verification times are complex functions of the
number of rows, number of columns, number of permutation arguments, number of
lookup tables, and the highest degree of the custom gate. Modeling the
computational burden for a given circuit is difficult as a result.

\section{\sn Architecture}
\label{sec:arch}

\sn is a library which takes as input an attested image and a high-level
specification of image transformations. It outputs the transformed image one or
more ZK-SNARK proofs attesting to the validity of the image transformations. \sn
further allows applications developers to specify new image transformations.

One critical requirement is the ability to specify \emph{multiple} image
transformations. Unfortunately, there is no work that allows multiple image
transformations to be securely \emph{and} privately applied in sequence. We
describe how to accomplish this via \sn in Section~\ref{sec:multi-image}.

At a high level, \sn operates by determining the maximum number of image
transformations it can place in a single ZK-SNARK proof, given the hardware
resources available. It then splits the chain of transformations based on these
constraints and produces a ZK-SNARK proof for each subset of transformations. As
we describe below, the verification key contains information about the
transformation. This verification key is specific to the transformation but the
generation can be amortized for many common transformations (resizing to
standard sizes, color space conversions, blurring filters, etc.).

To do so, \sn contains two components: a high-level optimizer and cost-modeler
for placing transformations, a transpiler for arithmetizing image
transformations, and an execution engine to produce the individual ZK-SNARKs. We
describe the components in more detail below.

We now describe \sn's security model and its limitations.

\minihead{Security model}
\sn assumes the existence of tamper-resistant attested cameras. Namely, we
assume these cameras produce attested images that are signed with a digital
signature. We further assume that the cameras cannot be tampered with (i.e.,
cannot be made to produce images that were not taken by the camera but signed).

Given the existence of tamper-resistant attested cameras, \sn \emph{only}
assumes the standard ZK-SNARK security model \cite{bunz2020transparent} and a
collision-resistant hash function \cite{rogaway2004cryptographic}. Informally,
the ZK-SNARK security assumptions state that the prover (the producer of the
image) and verifier (consumers of the media) interact only through the ZK-SNARK
and that adversaries (the producer of the image) is computationally bounded.

Under these standard security assumptions, \sn inherits the zero-knowledge,
non-interactive, knowledge-soundness, and completeness properties. In
particular, \sn will produce valid proofs of correctly applied transformations
and adversaries will not be able to produce invalid proofs.

While \sn preserves the zero-knowledge property of the inputs and
transformations, and completeness, there are other ways to attack media
consumers that are outside of the scope of this work. We describe these below.

\minihead{Limitations}
As described, \sn only preserves the zero-knowledge property of the input image
and intermediate transformations in the ZK-SNARKs. In particular, the revealed
output image may leak sensitive information if the transformations did not
properly hide the sensitive parts of the image. Such leakage is orthogonal to
this work.

Furthermore, \sn can only attest that the transformations of an attested image
were applied properly. Adversaries can manipulate the physical surroundings to
produce images that are attested but misleading. These adversaries could also
take an image of a deepfake. Validating such imagery is also orthogonal to this
work. However, we believe that using attested 3D depth information could help
mitigate these issues, such as LIDAR sensor data.

\section{Efficiently Proving Single Image Transformations}
\label{sec:single-image}

We first describe how \sn can produce ZK-SNARK proofs for single image
transformations. Although ZK-SNARKs allow arbitrary computations to be proven,
we focus on common image transformations that perform ``physical''
transformations (e.g., crops, rotations) or are the result localized pixel
computations (e.g., blurring, colorspace conversions, white balance adjustment).

One key requirement for \sn is to allow application developers to add additional
transformations. In order to enable extensibility, \sn uses the ZK-SNARK
verification key to specify the computation and the proof to verify that the
computation was done correctly. As a result, application developers do not need
to develop proving arguments that are specific to transformations and can
instead use the primitives \sn provides.

To preserve the privacy of intermediate results for transformations, \sn will
hash the input and output and reveal the hashes. Under standard cryptographic
assumptions of collision-resistant hashes, revealing hashes will reveal nothing
about the input data \cite{rogaway2004cryptographic}. \sn will only reveal the
final image. As we describe below, this enables optimizations for many
transformations. We describe \sn's full procedure for chaining transformations
in Section~\ref{sec:multi-image}.

In the remainder of this section, we describe how \sn uses the Plonkish
arithmetization to express common operations and how to implement specific image
transformations in \sn.

\begin{table}
\centering
\begin{tabularx}{\columnwidth}{l|X}
  Transformation & Description \\
  \hline
  \hline
  Crop      & Extract rectangular sub-image \\
  Rotate    & Rotate image \\
  Flip      & Flip image across x- or y-axis \\
  Translate & Move image across x- or y-axis \\
  Resize    & Change image resolution \\
  Censoring & Selectively black out pixels \\
  \hline
  RGB to YCbCr & Color space conversion \\
  YCbCr to RGB & Color space conversion \\
  \hline
  White balance & Adjust white balance \\
  Contrast & Adjust contrast \\
  \hline
  Sharpen & Apply sharpen filter \\
  Blur & Apply blur filter
\end{tabularx}
\caption{Summary of image transformations currently implemented in \sn. These
transformations are composable. Furthermore, other transformations can be
implemented in \sn.}
\label{table:transform-table}
\end{table}

\subsection{Common Operations}

We now describe how to perform copying, division, and dot products in the
Plonkish arithmetization. These operations are used across several image
transformations as building blocks.

\minihead{Copying}
One common operation is to copy the value from one cell to another. The Plonkish
arithmetization allows copying of cells to be done via the permutation argument.
This is easily specified in \halo. Importantly, the set of copied cells (which
are constrained to be equal to each other) can be extracted from the
verification key.

\minihead{Division}
Another common operation is integer division with a known, positive divisor.
Division is required to do fixed point arithmetic, which approximates floating
point arithmetic. As mentioned, the Plonkish arithmetization performs all
operations in a large finite field, which does not contain a native division
operation.

In order to represent division, we first consider the case of non-negative
dividend. Let $b = \lfloor \frac{c}{a} \rfloor$. Then, we have that
\[
c = b \cdot a + r
\]
where $0 \leq r < a$ is the integral remainder. Since the divisor $a$ is known
ahead of time, we can constrain the division with three cells and a lookup
constraint. Namely, we lay out three cells with values $c$, $b$, and $r$ with
the constraint that $c = b \cdot a + r$. We further constrain that $r$ be in the
range $0, ..., a - 1$ and $b$ to be in a valid range (the valid range of $b$
depends on the operation and can be computed ahead of time).

\minihead{Dot products}
The last common operation we describe are dot products. Dot products are used
for applying convolutional filters as image transformations and also for
pixel-level adjustments.

In particular, given vectors $\vec{a}$ and $\vec{b}$, we wish to constrain that
$c = \vec{a} \cdot \vec{b}$.  To do so, we can simply have the constraint that
\[
c = a_1 \cdot b_1 + \ldots + a_n \cdot b_n
\]
and assign $c, a_i, b_i$ to $2n+1$ cells. If one of the vectors is known ahead
of time (e.g., a fixed filter), the values can be omitted and instead hard-coded
into the constraint. This reduces the number of assigned cells to $n+1$ at the
cost of adding an additional selector.

\minihead{Hashes}
The \halo library contains several hash function implementations, which we use.
In this work, we assume that the camera produces the Poseidon hash
\cite{grassi2019poseidon} when the pixels are packed into the field element. As
we show, the hashing is by far the dominant cost of our circuits.

\subsection{Example Transformations}

Given the building blocks we have described, we describe how to use them to
construct common image transformations. For brevity, we describe a
representative sample of image transformations. The other image transformations
in Table~\ref{table:transform-table} can be implemented similarly.

\minihead{Crop}
An image crop takes an image and returns a rectangular subset of the image. Crops
can be used to remove sensitive information.

Since a cropped image is a subset of the input image, it can be done entirely
through copies. In particular, from an application developer perspective, a crop
is equivalent to copying a fixed set of cells. From a verifier perspective, a
crop is defined by the set of constraints on the cells.

Image translations, rotations, flips, and nearest neighbor resizing can be done
similarly.

\minihead{Selective removal}
A selective removal transformation selectively fills in a fixed set of pixels
with black pixels. The fixed set of pixels is typically a rectangle (e.g., a
portion of a document) or oval (e.g., a person's face). The application
developer may choose which patterns are permissible. This transformation can be
used to censor sensitive parts of the image, such as people's faces, signs, or
documents.

Similar to cropping, the selective removal transformation copies the
non-censored pixels from the original image. The censored pixels are copied from
a revealed cell that contains the value 0.

\minihead{Blur}
An image blur takes an image and returns an image with a subset of the image
with the blur filter applied. Similarly to crops, blurs can be used to remove
sensitive information. We describe how to blur the whole image (which is the
most resource intensive version), but any subset of the image can be blurred by
combining the blurred pixels with copied pixels from the original image.

One common method of applying a blur is to apply a standard Gaussian blur
filter, which is a convolutional filter, over the image
\cite{shapiro2001computer}. A convolutional filter is a local filter in which
pixels nearby the target pixel are combined in a fixed way. Namely,
\[
y[m, n] = \sum_{i = -c}^{c} \sum_{j = -c}^{c} x[m - i][n - j] \cdot h[i][j].
\]
for a convolutional filter $h$ of size $(2c+1) \times (2c+1)$.

To implement the Gaussian blur filter, we perform the following operation per
pixel. We perform the unrolled convolution as a dot product and divide the
result by the fixed point scalar. The result of the division is the clamped to
the valid image range (0 to 255). 

Other filters, such as a sharpening filter, can be implemented similarly.

\minihead{Contrast}
A contrast adjustment transformation takes an image and increases or decreases
the contrast of the image. Adjusting the contrast of the image can highlight
details that may be difficult to see in the unadjusted image.

One common way to adjust the contrast is to adjust each sub-pixel value $p$ by
some factor $f \in \mathbb{R}^+$ in the following way:
\[
p' = \textrm{RoundAndClip}(128 + f \cdot (p - 128), 0, 255)
\]
where RoundAndClip rounds the argument to the nearest integer and clips to 0 to
255 (the standard \texttt{uint8} range).

To implement the contrast adjustment, we can use a single lookup table. In
particular, $p$ takes 256 possible values. Since $f$ is known ahead of time, we
can encode the mapping $p \to p'$ with a lookup table.

Other transformations, such as white balance adjustments (which is also a
per-sub-pixel map), can be done similarly. Transformations that require
full-pixel computations may require additional computation, such as divisions or
other constraints.


\minihead{Color space conversion (RGB to YCbCr)}
A color space conversion transforms an image from one color space to another.
These conversions are often done before applying an image transformation that is
more amenable to the converted colorspace. For example, a luminance adjustment
requires only adjusting the Y value in the YCbCr colorspace, but requires
complex pixel computations in the RGB colorspace. We describe how to convert RGB
to YCbCr.

Given the sub-pixel R, G, and B values in RGB color space, we can compute the Y,
Cb, and Cr sub-pixel values in YCbCr as follows \cite{hamilton2004jpeg}:
{
\small
\begin{align*}
  Y  &=   0 &+ (0.299 \cdot    &R) &+ (0.587 \cdot    &G  &+ (0.114 \cdot    &B) \\
  Cb &= 128 &- (0.168736 \cdot &R) &- (0.331264 \cdot &G) &+ (0.5 \cdot      &B) \\
  Cr &= 128 &+ (0.5 \cdot      &R) &- (0.418688 \cdot &G) &- (0.081312 \cdot &B)
\end{align*}
}
in floating point. We can convert the floating point operations to fixed point
by multiplying all values by a scalar and rounding to the nearest integer. The
resulting value can be cast back to the valid range using division as described
above. The Y, Cb, and Cr values are rounded to the nearest integer.  They need
not be clamped due to the range of admissible values.

There are several choices of possible implementations for the RGB to YCbCr color
space conversion. One method would be to have a lookup table for the full pixel
values. However, this requires a lookup table of size $2^{24}$, which would
force the grid to have at least $2^{25}$ rows, as several additional rows are
required to ensure the zero-knowledge property. Requiring so many rows is
resource intensive. Nonetheless, it requires fewer columns (two per pixel).

In many circumstances, it is too resource intensive to have $2^{25}$ rows. As
such, we can implement the RGB to YCbCr color space conversion by explicitly
computing the linear functions. Each sub-pixel in YCbCr can be computed as a dot
product with known constants and a linear offset (0 or 128). As described above,
we can approximate the floating point arithmetic with fixed point arithmetic.
Finally, we can divide the results of the fixed point arithmetic by the
exponent. Clamping is not necessary as the domain constrains the range to be in
0 to 255.

Other colorspace conversions can be implemented similarly.

\minihead{Other transformations}
As shown in Table~\ref{table:transform-table}, we implement several other
transformations. These transformations largely follow the pattern of selectively
copying pixels and local transformations. Our API allows these patterns to be
implemented so they can be easily added and composed.

\subsection{Optimizations}

As described in Section~\ref{sec:snarks}, there are several important
considerations in the performance of the \halo proving system and Plonkish
arithmetization. We describe several optimizations to improve the performance of
the arithmetic circuits. Several similar optimizations were explored in the
context of executing machine learning models \cite{kang2022scaling}.

\minihead{Circuit size and packing rows}
One of the most critical factors to performance is circuit size. Circuit size is
determined by the number of rows and columns in the circuit. As we described in
Section~\ref{sec:snarks}, the number of rows must be a power of two.
Importantly, for circuit layout, there are sharp phase transitions in circuit
size. Furthermore, the memory pressure substantially increases with each power
of two.

Naively laying out our circuits would place one operation per row. However, this
can be memory inefficient. Namely, for a fixed number of cells, a circuit with
fewer rows and more columns is generally more memory efficient.

As such, \sn will pack as many operations per row as possible. In particular,
since the image transformations are largely local pixel computations, \sn will
lay out per-pixel computations and repeat them as many times as possible per
row.

For a fixed number of rows, \sn will optimize the column count by choosing the
minimum number of columns needed for the transformations and hashes. \sn will
similarly optimize the number of rows for a fixed number of columns. There is a
fixed minimum number of operations required for the hashes, which constrains the
minimum number of rows. \sn will estimate the total cost of various
configurations of rows and columns and pick the minimum cost plan under the
memory constraints.

\minihead{Hard-coding transformation-specific operations}
As described above, \sn provides the verification key in addition to the proof
certificate. The verification key encodes the transformations on a
per-transformation basis, so application developers need not construct custom
arguments for all possible transformation parameters.

In addition to relieving developer effort, per-transformation circuits allow \sn
to perform optimizations. In particular, wherever possible, \sn will avoid
copying values into new cells and will hard-code as many values as possible into
custom gates.

As an example of avoiding copies to new cells, consider the operation of a crop,
which extracts a subset of an image. Instead of copying the values into new
cells in the crop, \sn can instead return the subset of cells from the input.
These cells can subsequently be transformed, hashed, or revealed. This
optimization can make intermediate transformations \emph{free} in terms of
prover and verifier cost.

As an example of hard-coding transformation-specific values, consider the blur
operation. Instead of laying out the filter values, they can be directly encoded
in the custom gate as the constraint. This can save nearly 2$\times$ the number
of cells, greatly reducing the computational burden.

\minihead{Sharing lookup tables}
A number of operations require lookup tables. For example, they may be used to
perform division or clamp a value.

Wherever possible, \sn will reuse lookup tables. In particular, a common
operation is a clamping operation, which performs
\[
\textrm{Clamp}(x) = \textrm{Max}(\textrm{Min}(x, 255), 0).
\]
Given the range of possible values of $x$, we can construct two lookup tables
for the input and output respectively. These lookup tables can be shared across
transformations that require the clamping operator.

Similarly, \sn can reuse division remainder range checks for operations that use
the same scale factor.

\begin{figure}
  \centering
  \begin{subfigure}[h]{\columnwidth}
    \includegraphics[width=\textwidth]{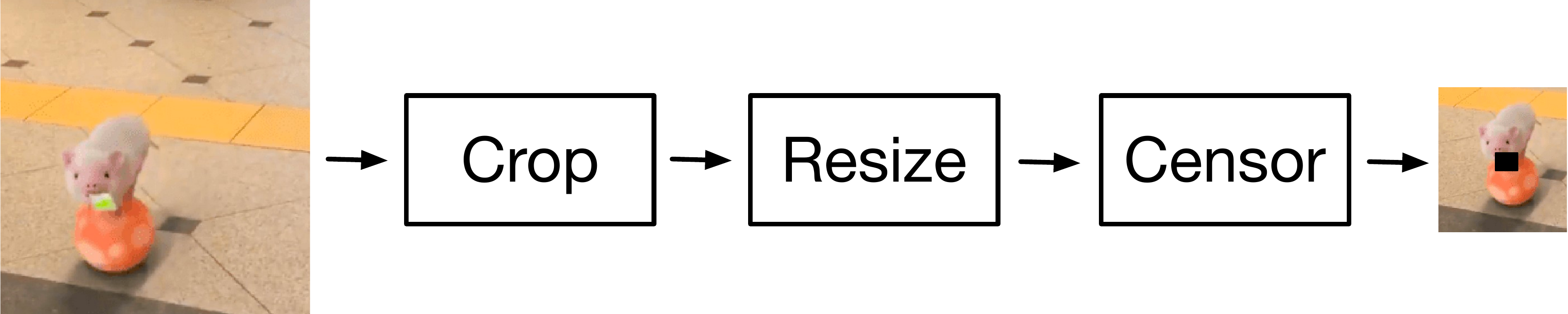}
    \caption{Example of a series of image transformations which hides sensitive
    information (the card in the pig's mouth).}
  \end{subfigure}
  \vspace{0.5em}
  \begin{subfigure}[h]{\columnwidth}
    \includegraphics[width=\textwidth]{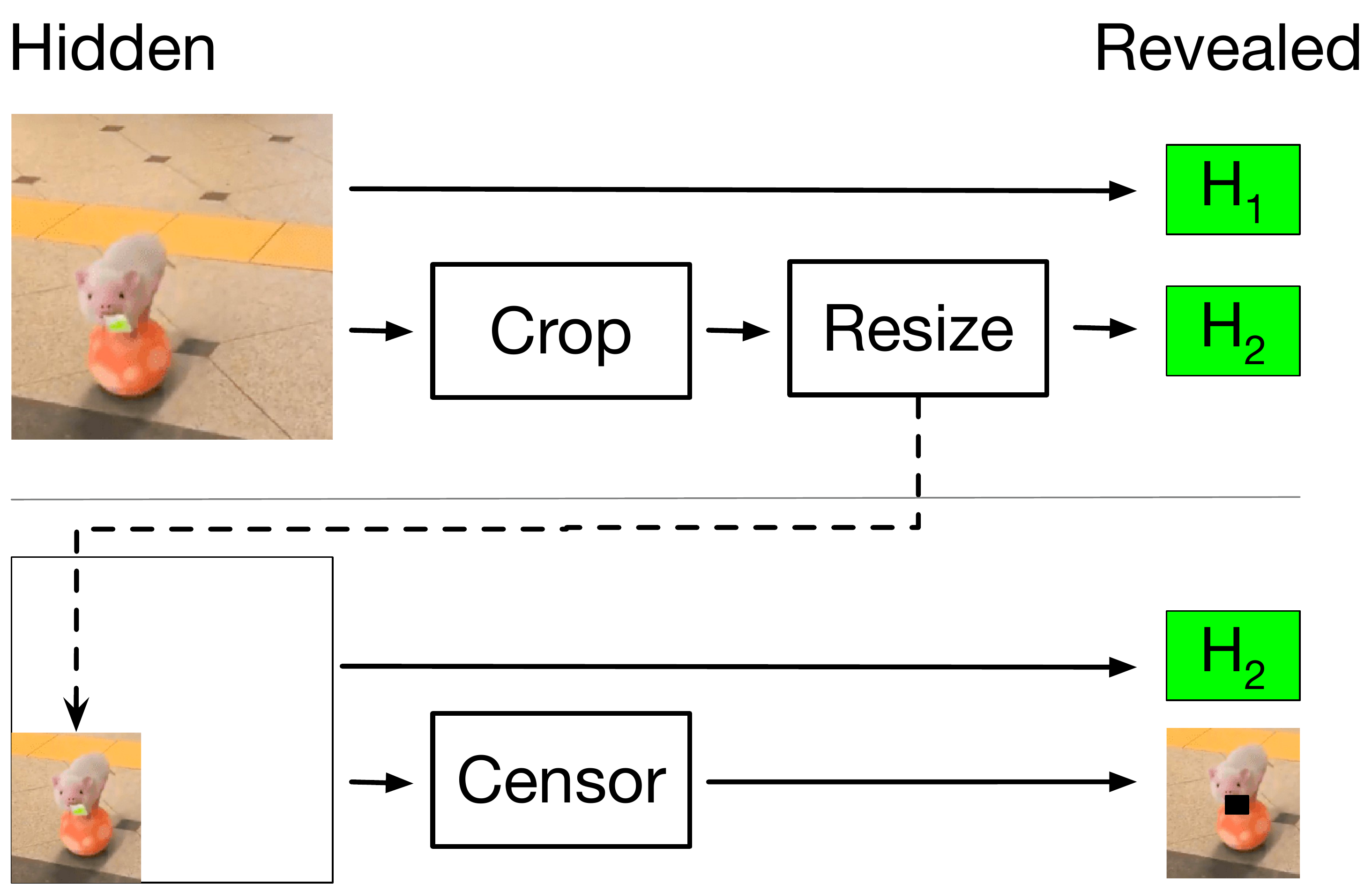}
    \caption{Example of splitting the image transformations across multiple
    ZK-SNARKs while preserving intermediate transform privacy. Only the hashes
    ($H_1, H_2$) and final image are revealed.}
  \end{subfigure}
  \caption{Schematic of \sn's procedure to split transformations across
  ZK-SNARKs while hiding intermediate images.}
  \label{fig:multi-transform}
\end{figure}

\section{Securely Chaining Image Transformations}
\label{sec:multi-image}

In many circumstances, the initially attested image is edited with several
transformations before being released. For example, a typical series of
transformations may include a crop and censoring (to hide sensitive
information), white balance adjustment, sharpen, and contrast adjustment (for
increased legibility), and a resize (to transform the image to a standard size).

One such way to chain a series of image transformations would be to do all the
transformations in a single circuit. Unfortunately, this increases the
computational burden of creating the ZK-SNARK proof. In particular, increasing
the circuit size increases the memory burden of proving, which can result in
insufficient memory resources.

Another way to chain transformations would be to ZK-SNARK individual transforms
and release the intermediate images. Unfortunately, revealing the intermediate
images compromises the privacy guarantees of the ZK-SNARKs (otherwise, the
initial image could have been released).

\begin{table*}[t!]
\centering
\begin{tabular}{llllll}
  Transformation & Key generation & Proving & Verification & Proof size & Peak memory usage \\
  \hline
  Crop (HD $\to$ SD)   & 5.63s &  7.8s & 6.10ms & 7040 bytes  & 1.72 GB \\ 
  Resize (HD $\to$ SD) & 5.60s &  7.5s & 5.84ms & 7040 bytes  & 1.72 GB \\
  Contrast             & 14.4s & 14.3s & 5.57ms & 5088 bytes  & 4.08 GB \\
  White balance        & 16.1s & 15.7s & 5.76ms & 5792 bytes  & 4.63 GB \\
  RGB2YCbCr            & 31.5s & 65.9s & 9.39ms & 14592 bytes & 11.0 GB \\
  YCbCr2RGB            & 31.9s & 67.7s & 10.1ms & 14592 bytes & 11.0 GB \\
  Convolution          & 69.8s & 81.7s & 8.14ms & 11200 bytes & 15.74G
\end{tabular}
\caption{Performance measurements for \sn's image transformation
implementations. Our implementations are efficient: they can take as few as 5.6
ms to verify, as few as 7.5s to prove, and require a peak memory usage of at most
15.7GB. As shown, our transformations can be done on commodity laptops.}
\label{table:transformation-benchmarks}
\end{table*}

To address this issue, \sn will split image transformations across ZK-SNARK
proofs. However, instead of releasing the intermediate images, \sn will exhibit
a private witness (i.e., the pixel values in the intermediate image) and hash
the intermediate image in the ZK-SNARK. \sn will only reveal the hashes of the
intermediate images, not the pixel values themselves. We show a schematic of
this process in Figure~\ref{fig:multi-transform}.

This raises a natural question: how should \sn pack transformations between
ZK-SNARKs given limited memory resources? Every extra ZK-SNARK proof adds an
additional hash operation, which is expensive. As such, \sn aims to reduce the
total number of ZK-SNARK proofs. \sn will do this by cost modeling the
transformations and packing as many transformations as possible per ZK-SNARK
proof subject to the memory limit. 

\section{Evaluation}
\label{sec:eval}

In this Section, we evaluate \sn on a range of image transformations and
settings. We show that \sn can produce proofs that can be verified in as little
as 5.6ms on HD image transformations. We further show that \sn's image
transformation implementations are efficient, taking as few as 7.5s to prove.
When implemented end-to-end (which no other work has done), \sn can prove
transformations on commodity hardware with as little as \$0.48 per image.

\subsection{Evaluation Setup}
We evaluated \sn on one or more image transformations, both end-to-end (i.e.,
secured by hashing the inputs/outputs) and the transformations themselves. For
consistent results, we use the Amazon Web Services (AWS) \texttt{r6i.16xlarge}
instance, which has 64 vCPU cores and 512 GB of memory. While large, widely
available desktop devices (e.g., the Mac Pro) has similar amounts of resources.
We defer a full discussion of hardware limitations to
Section~\ref{sec:eval-discussion}.

We measured the time to generate the ZK-SNARK proof (i.e., the time for the
producer of the image to prove the transformations), the verification time
(i.e., the time for the image consumer to verify the proof), the proof size, and
peak memory usage (using \texttt{heaptrack}). Where applicable, we further
estimate dollar costs of the proving time using AWS spot instances.

To compute the hashes of inputs and outputs, we use the Poseidon hash function
\cite{grassi2021poseidon}, which can be implemented efficiently in ZK-SNARKs. We
use a width of 3 rate of 2, which gives 128 bits of security. As we show,
computing the hashes is by far the dominant cost of end-to-end image
transformations.

\subsection{Evaluating Transformations}

\minihead{\sn performance}
We first benchmark our implementations of single image transformations. Several
of our transformations are implemented logically when combined with other
transformations (crops, rotations, censoring, etc.). For these transformations,
we benchmarked the versions that copied the values from the input.

Our implementation of the transformations in \sn can take as little as 7.8
seconds to prove and 5.6 ms to verify
(Table~\ref{table:transformation-benchmarks}). Our most expensive transformation
(the convolution) takes 82 seconds to prove and 8.1 ms to verify. The maximum
peak memory usage across transformations is 15.7 GB.

As these results show, \sn's implementations of image transformations are
efficient and can transform HD images on widely available laptops (e.g., the
MacBook Pro). These results demonstrate the feasibility of verifying high
quality images.

\minihead{Comparison to PhotoProof}
The closest work we are aware of is PhotoProof \cite{naveh2016photoproof}.
PhotoProof verifies the signatures and hashes outside of the ZK-SNARKs, which
can be done if the intermediate images are revealed. The implementation of
PhotoProof was not available at the time of writing and authors did not respond
to our correspondences to make their code available.

To compare against PhotoProof, which applies a single image transformation at a
time, we executed \sn on contrast, the most expensive operation PhotoProof
considered. We further matched hardware as closely as possible by using the
\texttt{r6i.xlarge} instance, which had 4 vCPU cores (half as many threads as in
the evaluation of \cite{naveh2016photoproof}) and 32 GB of RAM. We used
128$\times$128 images as in PhotoProof.

PhotoProof requires 306 seconds to prove (ignoring the cost of key generation)
and 500ms to verify transformations on 128$\times$128 images. In contrast, \sn
takes 2.74 seconds to prove (\emph{including} the cost of key generation) and
5.3ms to verify, which corresponds to a 112$\times$ and 94$\times$ speedup,
respectively.

\begin{table}[t!]
\centering
\begin{tabular}{llr}
  Operation     & \sn & No privacy \\
  \hline
  Contrast      & 6.27ms & 9.82ms \\
  White Balance & 6.22ms & 8.34ms \\
  RGB2YCbCr     & 6.50ms & 13.24ms \\
  YCbCr2RGB     & 5.44ms & 13.04ms \\
  Convolution   & 4.69ms & 78.25ms
\end{tabular}
\caption{Verification time of \sn with privacy and a no-privacy alternative for
various image transformations. As shown, \sn outperforms on verifying images.}
\label{table:no-privacy}
\end{table}

\begin{table*}[ht!]
\centering
\begin{tabular}{llllll}
  Transformation & Key generation & Proving & Verification & Proof size & Peak memory usage \\
  \hline
  Crop (HD $\to$ SD)   & 246.1s & 328.2s & 6.90ms & 3040 bytes &  70.7 GB \\
  Resize (HD $\to$ SD) & 246.6s & 328.2s & 5.33ms & 3040 bytes &  70.7 GB \\
  Contrast             & 283.1s & 363.8s & 6.27ms & 3712 bytes &  85.3 GB \\
  White balance        & 293.9s & 359.6s & 6.22ms & 3776 bytes &  87.9 GB \\
  RGB2YCbCr            & 276.3s & 617.5s & 7.96ms & 6496 bytes &  89.1 GB \\
  YCbCr2RGB            & 276.2s & 606.9s & 8.02ms & 6496 bytes &  89.1 GB \\
  Convolution          & 339.5s & 428.5s & 4.69ms & 4672 bytes & 102.2 GB
\end{tabular}
\caption{Performance measurements of \sn's image transformations when including
the hash of the input. Despite requiring more computational resources
for proving, the proofs can be verified in as little as 4.7 ms and at most 8.0 ms.
Unfortunately, the hash can take up to 95\% overhead for proving, requiring
dramatically more time and memory.}
\label{table:transformation-e2e}
\end{table*}

\minihead{Comparison to no privacy alternative}
We then asked how \sn's performance compared to the no privacy alternative.
Namely, the owner of the image releases the attested image publicly and the
verify performs the transformations directly.

We implemented the transformations in Rust and benchmarked the time to perform
the transformations. As shown in Table~\ref{table:no-privacy}, the cost of
performing the transformations directly can actually be greater than the cost of
verifying the transformations. Although surprising, ZK-SNARKs can provide
sub-linear verification of computations.

\subsection{Evaluating End-to-End Transformations}

We then benchmarked \sn on end-to-end image transformations, when including the
hash of the inputs. This can be used to keep the privacy of the input data.

As shown in Table~\ref{table:transformation-e2e}, the verification times are
still minimal: at most 8.0 ms. Unfortunately, the hashing can be up to
20$\times$ slower than the image transformation itself for proving. This can
result in proving times up to 617 s and peak memory usage up to 102 GB.
Nonetheless, these operations can be run on powerful, commercially available
desktops (e.g., the Mac Pro). Furthermore, these transformations can cost as
little as \$ 0.28 on cloud hardware.

\begin{table*}[ht!]
\centering
\begin{tabular}{llllll}
  Transformation & Key generation & Proving & Verification & Proof size & Peak memory usage \\
  \hline
  Crop (HD $\to$ SD)   & 432.6s &  557.1s & 6.22ms &  6112 bytes & 139.1 GB \\
  Resize (HD $\to$ SD) & 431.8s &  556.8s & 5.62ms &  6112 bytes & 139.1 GB \\
  Contrast             & 823.8s & 1029.1s & 8.16ms & 12608 bytes & 284.3 GB \\
  White balance        & 839.4s & 1027.9s & 8.60ms & 12672 bytes & 287.1 GB \\
  RGB2YCbCr            & 816.1s & 2198.0s & 15.4ms & 26144 bytes & 307.9 GB \\
  YCbCr2RGB            & 815.5s & 2236.2s & 14.8ms & 26144 bytes & 307.9 GB \\
  Convolution          & 897.9s & 1300.3s & 9.32ms & 15232 bytes & 305.3 GB
\end{tabular}
\caption{Performance measurements of \sn's image transformations when including
the hash of the input and output. Despite requiring more computational resources
for proving, the proofs can be verified in as little as 5.6ms and at most 15.4ms.
Unfortunately, the hash can take up to 95\% overhead for proving, requiring
dramatically more time and memory.}
\label{table:transformation-hidden}
\end{table*}

\subsection{Evaluating End-to-End Hidden Images}

We then benchmarked \sn on end-to-end image transformations, when including
hashes of the inputs and outputs. This can be used for producing attestations to
hidden images. Namely, for each operation, we measured the performance when
hashing the input and output.

As shown in Table~\ref{table:transformation-hidden}, the verification times are
still minimal: at most 15.4 ms (a sublinear increase compared to a single hash).
As with input privacy preserving transformations, hashing the output can
dramatically increase the proving overheads of image transformations. This can
result in proving times up to 2236s and peak memory usage up to 308 GB.
Nonetheless, these operations can be run on powerful, commercially available
desktops (e.g., the Mac Pro) as before. Furthermore, these transformations can
cost as little as \$0.48 on cloud hardware.

The cost of some operations (crop, resize) are substantially cheaper than the
cost of others. This is because for both the crop and resize, the output image
is SD (720$\times$480), which contains 2.7$\times$ fewer pixels compared to HD
images. Since hashing is the predominant cost, this results in substantially
cheaper output hashes, which decreases the key generation time, proving time,
and peak memory usage.

\begin{table}
\centering
\begin{tabular}{llll}
  Operation & \sn & \sn  & No privacy \\
            &     & + boilerplate \\
  \hline
  Crop          & 20  & 52  & 12 \\
  Resize        & 39  & 74  & 18 \\
  Contrast      & 112 & 132 & 9 \\
  White balance & 135 & 157 & 8 \\
  RGB2YCbCr     & 192 & 212 & 24 \\
  YCbCr2RGB     & 212 & 232 & 26 \\
  Convolution   & 174 & 195 & 23
\end{tabular}
\caption{Lines of code to implement \sn's transformations and the no-privacy
versions. Even the most complex transformation requires at most 232 lines of
code.}
\label{table:loc}
\end{table}

\subsection{Developer Effort}

We asked how much developer overhead \sn requires. To measure this, we compared
the number of lines of code that implementing transformations in \sn required
compared to manually implementing the transformations without privacy in Rust.
We measured the lines of code for the transformations only in \sn,
transformations with boilerplate code in \sn, and our no-privacy, expert-coded
transformations in Rust. The boilerplate code in \sn is required to interface
with the \halo library.

We show results in Table~\ref{table:loc}. As shown, \sn's transformations can be
written in as few as 20 lines of code. The developer effort increases with the
complexity of the transformation, but remains manageable, under 232 lines of
code in all settings.

\subsection{Discussion of Results}
\label{sec:eval-discussion}

As we have shown, \sn is able to transform images efficiently, with peak memory
usage of 15.7 GB and verification times as low as 5.9ms. However, when including
the hashes of the inputs and outputs, the peak memory usage increases to 305
GB. Although feasible on a powerful desktop, this limits the feasibility of
deploying \sn widely. In particular, there are modern image editing methods
which leverage deep neural networks (that no prior work addresses) that are
currently computationally expensive, although possible to implement in \sn.

Fortunately, there are several ways to improve the performance of \sn; we
highlight several such methods. First, proving systems are rapidly improving.
Recent advances in proving systems can asymptotically reduce the memory usage
from $O(n \log n)$ to $O(n)$ \cite{chen2022hyperplonk} while simultaneously
improving proving times. Second, ZK-SNARK-friendly hash functions are rapidly
being developed. Since the hashing operations take up to 95\% of the
computational resources, any developments in improved ZK-SNARK-friendly hash
functions will also dramatically improve \sn. Third, ZK-SNARKs are amenable to
hardware acceleration \cite{xavier2022pipemsm, lu2022cuzk}. We hope that
improvements in ZK-SNARK technology will unlock widespread availability of
verified images and that \sn can serve as a platform for such work.

\section{Related Work}
\label{sec:rel-work}

\minihead{Attested cameras}
Attested cameras digitally sign images immediately on capture to ensure the
validity of an image \cite{friedman1993trustworthy}. These cameras can be
purchased on the market today, such as Sony's Alpha 7 IV camera
\cite{sony2022camera}. Although they can sign the images on capture, they cannot
attest to image transformations of the initially captured image. In this work,
we focus on the problem of attesting to downstream image transformations.

\minihead{Provable image transformations}
Several prior research projects have designed systems for verified image
transformations \cite{naveh2016photoproof, ko2021efficient, datta2022using}.
However, this work has three major drawbacks as we have described: they require
\emph{the original image to be revealed} \cite{naveh2016photoproof,
datta2022using}, only operates on impractically small images (128$\times$128)
\cite{naveh2016photoproof, datta2022using}, or require specific cryptographic
arguments for new image transformations \cite{naveh2016photoproof,
ko2021efficient}. In this work, we scale zero-knowledge image transformations to
HD images, provide a secure method of chaining together arbitrary image
transformations arbitrarily many times.

\minihead{Other cryptographic primitives}
There are a range of other cryptographic primitives that can provide privacy or
security. Two popular methods are multiparty computation (MPC) and homomorphic
encryption (HE). MPC allows multiple parties to compute a function without
revealing information about the inputs or intermediate states
\cite{goldreich1998secure}. Unfortunately, MPC requires that all parties be
online during the duration of the computation. As a result, it is not suitable
for our setting. HE allows parties to perform computation over encrypted data
without first decrypting the data \cite{armknecht2015guide}. Unfortunately, HE
is incredibly expensive. In our setting, this would place the burden of
computation on the consumers of the images, which may be consumer devices such
as laptops or phones. As such, it is infeasible to use HE in our setting.

\minihead{Deepfake detection}
Aside from using cryptographic primitives, the literature has proposed detecting
deepfakes to combat them \cite{lyu2020deepfake}. There are a range of benchmarks
\cite{dolhansky2020deepfake, zi2020wilddeepfake} and techniques proposed to
detect deepfakes \cite{coccomini2022combining, zhao2021multi,
guarnera2020deepfake}. While promising, there are a number of challenges with
these methods. As with other security problems, deepfakes are rapidly evolving,
with generation capabilities improving rapidly. Detection methods that
work now may not work in the future. Furthermore, adversarial examples
\cite{goodfellow2014explaining} can attack deepfake detection methods. In this
work, we propose instead to use ZK-SNARKs to attest that images were taken by a
particular camera instead, which bypasses such issues.

\section{Conclusion}
\label{sec:conclusion}

In this work, we present \sn, the first system for attesting to arbitrarily
transformed images at HD scale. \sn accomplishes this by producing ZK-SNARK
proofs of arbitrary image transforms and chaining sequences of proofs in a
secure manner. \sn is able to attest to HD images on commodity hardware,
producing proofs that take as little as 5.6 milliseconds to verify. Furthermore,
application developers are able to extend \sn with new image transformations. We
hope that \sn serves as a platform for research in attested images.

\section*{Acknowledgments}

This work is supported in part by the Open Philanthropy project.

%

\bibliographystyle{plainurl}
\bibliography{paper}

\end{document}